\documentclass[preprint]{acm_proc_article-sp}

\usepackage{url}
\usepackage{algorithm}

\usepackage{tabularx}
\usepackage{multirow}
\usepackage{hhline}

\usepackage{fixltx2e}

\usepackage{amsmath}

\usepackage{color}
\usepackage{graphicx, subfigure}

\begin{document}

%
\title{Expressing social attitudes in virtual agents for social training games}

\numberofauthors{5}

\author{
Nicolas Sabouret\\
\affaddr{LIMSI-CNRS, Univ. Paris-Sud}\\
\email{nicolas.sabouret@limsi.fr}
\and
Haza\"el Jones\\
\affaddr{Univ. Pierre \& Marie Curie}\\
\email{hazael.jones@lip6.fr}
\and
Magalie Ochs\\
\affaddr{Telecom ParisTech}\\
\email{magalie.ochs@telecom-paristech.fr}
\and
Mathieu Chollet\\
\affaddr{Telecom ParisTech}\\
\email{mathieu.chollet@telecom-paristech.fr}
\and
Catherine Pelachaud\\
\affaddr{Telecom ParisTech}\\
\email{catherine.pelachaud@telecom-paristech.fr}
}

\maketitle


\begin{abstract} 
The use of virtual agents in social coaching has increased rapidly in the last decade. In order to train the user in different situations than can occur in real life, the virtual agent should be able to express different social attitudes. In this paper, we propose a model of social attitudes that enables a virtual agent to reason on the appropriate social attitude to express during the interaction with a user given the course of the interaction, but also the emotions, mood and personality of the agent. Moreover, the model enables the virtual agent to display its social attitude through its non-verbal behaviour. The proposed model has been developed in the context of job interview simulation. The methodology used to develop such a model combined a theoretical and an empirical approach. Indeed, the model is based both on the literature in Human and Social Sciences on social attitudes but also on the analysis of an audiovisual corpus of job interviews and on post-hoc interviews with the recruiters on their expressed attitudes during the job interview.
\end{abstract}



%
%
\section{Introduction}

\begin{sloppypar}
Social coaching workshops, organized by inclusion associations, social
enterprises or coaching companies, constitute a common approach to
help people in acquiring and improving their social
competencies. However, it is an expensive and time-consuming approach
that relies on the availability of trained practitioners as well as
the willingness of the people to engage in exploring their social
strengths and weaknesses in front of their peers and practitioners.
\end{sloppypar}
The use of virtual agents in social coaching has increased rapidly in
the last decade. Projects such as those by Tartaro and Cassell
\cite{tartaro2008playing} or e-Circus \cite{aylett2009affective}
provide evidence that virtual agents can help humans improve their
social skills and, more generally, their emotional intelligence
\cite{goleman2006social}.

Most of these models of virtual agent in social coaching domain have focused on the simulation of emotions \cite{tartaro2008playing,aylett2009affective,hoque2013}. However, in social coaching domain, a virtual agent may embody different social roles, such as virtual recruiter in a job interview simulation. Given its role and the course of the interaction, the virtual agent should be able to express different social attitudes to train the user in different situations that can occur in real life. Thus, one of the key elements of a virtual agent in the domain of social coaching is then its capacity to reason on social attitudes and to express them through its behaviour. This is why, in this paper, we propose a model of social attitudes for expressive virtual agents.

Social attitude can be defined as ``\textit{an affective style that spontaneously develops or is strategically employed in the interaction with a person or a group of persons, colouring the interpersonal exchange in that situation (e.g. being polite, distant, cold, warm, supportive, contemptuous)}'' \cite{scherer2001appraisal}. As highlighted in \cite{Snyder1983,Wegener1994}, one's social attitude depends on one's personality but also one's moods that is directly influenced by the course of the interaction. One's social attitude is mainly conveyed by one's non-verbal behaviour \cite{Argyle88}. For instance, a dominant attitude is expressed through large and frequent gestures, a head directed upwards and larger amounts directed towards the interlocutor.  

In this paper, we propose a model of social attitudes that enables a virtual agent to adapt its social attitude during the interaction with a user. It considers user's emotions, mood and personality and compute and display agent's appropriate social attitudes. Moreover, the model enables the virtual agent to display its social attitude through its non-verbal behaviour. The proposed model has been developed in the context of job interview (JI) simulation. This context is interesting for several reasons: 1) social attitude plays a key role in JI: the applicant tries to fit the social norm and 2) the recruiter is in a role-play game, which can be simulated with a virtual agent. Moreover, job interview is a type of social coaching situation with high social impact. The methodology used to develop such a model combined a theoretical and an empirical approach. Indeed, the model is based both on the literature in Human and Social Sciences on social attitudes but also on the analysis of an audiovisual corpus of job interviews and on post-hoc interviews with the recruiters on their expressed attitudes during the interview. 

The paper is organized as follows. In the next section, we discuss existing works in affective computing and intelligent virtual agent for social coaching. Section 3 presents the general architecture including a cognitive module to reason on social attitudes and an expressive module for the display. Section 4 presents the corpus that was collected to design the model. The cognitive module to compute the social attitude to express is introduced Section 5. The expressive module to give the capability to the virtual agent to adapt its non-verbal behaviour to convey social attitude is presented Section 6. We conclude Section 7.

\section{Related work and Theoretical background}
\label{Relatedwork}

 A social attitude can be defined as ``\textit{an affective
style that spontaneously develops or is strategically employed
in the interaction with a person}''  \cite{scherer2001appraisal}.
Social attitudes can be described by categories, but also can be represented along 2 dimensions: friendliness and dominance as in the interpersonal circumplex introduced by Leary \cite{Leary1957}.
Several research has shown that one's social attitude is influenced by one's affective state (e.g. his emotions and moods \cite{forgas84}) and the course of the interaction (e.g. the affective reaction of the other \cite{Wegener1994}). For instance in \cite{Wegener1994}, Wegener et al. show that a positive mood can help influence a change of attitude in the interlocutor, and that people tend to feel a higher likelyhood toward interlocutors that are in a positive mood.

Our aim is to develop a model of social attitudes that can be used by virtual agents to select determine expressive behaviours given its simulated affective state and the course of the interaction. Several models have been proposed in the domain of affective computing to compute the virtual agent's emotions, moods and social relations considering different personality. More recently, several applications of intelligent virtual agents and affective models in the context of job interviews have been proposed \cite{hoque2013, Batrinca2013, tardis2013}. However, most of these models are based on a reactive approach that connect expressed attitudes to perceptions in a reactive manner. On the contrary, our model propose to build and reason on a representation of social attitudes, so as to develop strategic intentions in the context of job interview.

Several research in Human and Social Sciences has shown that most modalities of the body are involved when conveying attitudes: smiles can be signs of friendliness\cite{Burgoon84}, performing large gestures may be a sign of dominance, and 
a head directed upwards can be interpreted with a dominant attitude \cite{Carney05}. 
However, an attitude is not solely displayed by a sign. It is important to consider the succession of signs displayed by the agent as well as the signs displayed by the interlocutors. it is also crucial to consider how the signs of both interlocutors relate to each others. For example, it is only by looking at the sequencing of smile, gaze and head aversion that we can differentiate between amusement, shame and embarrassment, affects expressing different values of dominance \cite{Keltner95}.

Models of social attitude expression for virtual agents have already been proposed. For instance, in \cite{Ballin04}, postures corresponding to a given attitude were automatically generated for a dyad of agents. Ravenet \textit{et al.} \cite{Ravenet13} proposed a user-created corpus-based methodology for choosing the behaviours of an agent conveying an attitude along with a communicative intention.
The SEMAINE project used ECAs capable of mimicking and reacting to the system user's behaviour to convey a personality. Each of these works used either limited modalities, or a non-interactive (agent-agent) context. Also, none of these works looked at the sequencing of the agent's signal. In this article, we present model of social attitude expression that considers the sequencing of non-verbal behaviour (Section 6).

\section{General Architecture}

Our approach of social coaching is based on role playing games (or interview simulation) with post-interview coaching sessions with an expert. Our goal is to provide the users with a simulation platform that allow people to train against a virtual agent. The coaching aspect is not part of this paper: we focus on the simulation platform and more precisely on the human-virtual agent interaction. Our goal is to define an architecture that supports intelligent expression of social attitudes, based on a cognitive architecture and a set of non-verbal behaviour bricks.

In this model, the interview simulation for social coaching involves two main actors, namely the participant (\emph{i.e.} the person that is training on the system) and the interlocutor (\emph{i.e.} person or virtual agent that respond to the trainee). In our platform, the interlocutor is replaced by a virtual agent.

Although the model presented here is general and can be applied to different interaction situations, our corpus and the derived cognitive architecture and non-verbal behaviour are designed in the context of job interview simulations when the agent acts as recruiter.

\begin{figure}[h!]
\begin{center}
\includegraphics[width=\columnwidth]{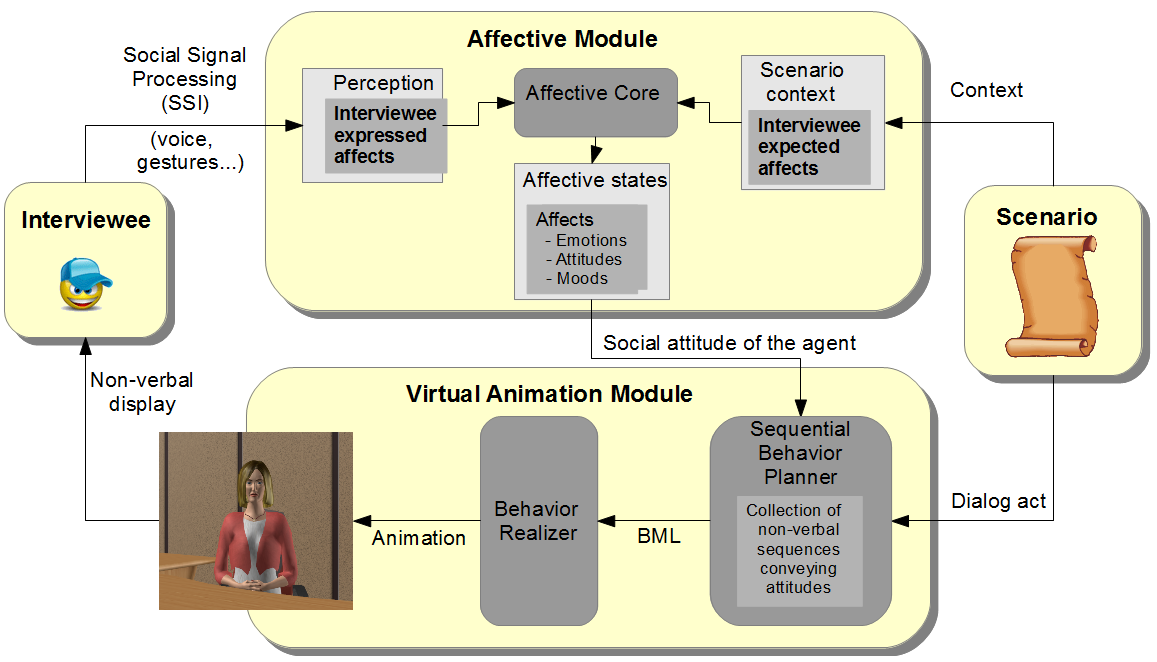}
\caption{General architecture}
\label{fig:general architecture}
\end{center}
\end{figure}

Fig.~\ref{fig:general architecture} presents our general architecture.

This architecture is organized as follows. First, a \textit{scenario} describes the context of the interaction and the expectations at the different stages of the interview. For example, after a tough question, the interviewer will expect negative affects from the interviewee (distress, agitation). 
These elements are used by the \textit{Affective Module}. The interaction loop starts from the recognition of affects expressed by the interviewee. Indeed, during a job interview, the emotions expressed by the interviewee constitute a key element influencing the recruiter's attitudes, the interviewee's emotions reflecting his social skills. In \cite{Sieverding2009}, a study shows that people who tried to suppress or hide negative emotions during a job interview are considered more competent by evaluators. Consequently, the \textit{Affective Module} takes as input the affective state of the interviewee. Several tools have been developed to obtain this information in real-time. In our work, we use the SSI framework \cite{wagner2013social} that recognizes the user's emotions expressed through his voice and his facial expressions.    

The \textit{Affective Module} (Section \ref{affectiveModel}) reasons both on the affects (of the user and the virtual agent) and on the context of the interaction to derive the virtual agent's social attitudes. These attitudes are turned into non-verbal behaviours in the virtual agent \textit{Animation Module} (Section \ref{attitudesDisplay}) and used in the next interaction step. The Animation Module is based on the SEMAINE platform \cite{Schroder12}, which includes an international common multi-modal behaviour generation framework. The \textit{Intent Planner} generates the communicative intentions (what the agent intents to communicate). This infomation is provided by the scenario. The social attitude to express is transmitted to the \textit{Behaviour Planner}. The \textit{Behaviour Planner} transforms these communicative intentions into a set of signals (e.g. speech, gestures, facial expressions). Moreover, the \textit{Behaviour Planner} has been extended to give the capability to the agent to display different social attitudes (Section 6). Finally, the \textit{Behaviour Realizer} outputs for each of these signals the animation parameters.   



In the following sections, we first present the corpus that was used to build our social attitude model. We explain how the corpus was collected, based on mock job-interviews, and it was annotated. We then give details of the Affective Module and the Animation Module and we show their interrelation.

\section{Corpus} 
\label{corpus}

We have collected a corpus of real interpersonal interaction between
professional recruiters and job seekers interviewees. The recordings consisted in creating a
situation of job interviews between 5 recruiters and 9 interviewees.
The setting was the same in all videos. The recruiter and the
interviewee sat on each side of a table. A single camera embracing the
whole scene recorded the dyad from the side. This resulted in a corpus
of 9 videos of job interview lasting between 15 and 20 minutes
each. We discarded 4 videos as the recruiter was not visible due to
bad position of the camera. Out of the 5 remaining videos, we have so
far annotated 3, for a total of 50 minutes and 10 seconds of video. 
%

The non-verbal behaviours of the recruiters during the job interview have been annotated. We also annotate information on the  interaction: the 
 \textit{turn taking} (i.e. who is speaking), the \textit{topic of the discussion} (i.e. document related or general), the perceived affects of the interviewee (e.g. embarassed, relieved, bored...),  and the attitude of the recruiter (i.e. the level of perceived dominance and friendliness of the recruiter). Different modalities of the recruiter's non-verbal behaviour have been annotated (e.g. the gaze behaviour, the gesture, the posture, the head movements, etc). The coding scheme, the resulting annotations, and the inter-annotators agreements are described in more details in \cite{Chollet13MMC}. These annotations have been used to construct the animation module for the virtual agent's expression of social attitudes (Section 6). Moreover, the annotation on the recruiter's social attitude has been used for the evaluation of the affective
model to check that it produces outputs that correspond to the human behaviour (Section~6).       


In addition to these videos, post-hoc interviews with the recruiters
were used to elicit knowledge about the expectations and mental states
during the interview, following the methodology proposed by
\cite{porayska2013}. This knowledge was used in the affective module to select the relevant social attitudes and to set up the rules to give the capability to the virtual agent to select the appropriate social attitude to express.

%


\section{Reasoning about attitudes}
\label{affectiveModel}
%

\subsection{Overview of the model}
The Affective Module is based on a set of rules that compute categories
of emotions, moods and attitudes for the virtual recruiter, based
on the contextual information given by the scenario and the detected
affects (emotions, moods and attitudes) of the participant (Section 3).  The computation of the virtual agent's emotions is based on the OCC model \cite{Ortony1988b} and the computation of the agent's moods is based on the ALMA model \cite{Gebhard2005}. The details of the computation of emotions
and moods will not be presented in this paper; it can be
found in \cite{jones2013}.

Formally, all affects in our model correspond to variables taking
values in the interval $[0,1]$ and we denote $\mathcal{A}$ the set of
all affects (moods, emotions and attitudes), with $\mathcal{A}^+$ the set of positive affects (joy,
focused, etc)  and $\mathcal{A}^-$ the set of negative ones (distress,
anxious, etc). Let $\mathcal{A}_d(t)\subset\mathcal{A}$ be the set of
affects that are detected ($d$) at a time $t$ of the interaction 
We separate them in three subsets $E_d(t)$ (emotions), $M_d(t)$ (moods)
and $A_d(t)$ (attitudes) and we denote $val(a)$ the value associated
to each detected affect $a$. Similarly, we receive from the scenario a
set of expected ($e$) affects $\mathcal{A}_e(t)\subset\mathcal{A}$. Expected affects are
the expectations of the interviewer about the interviewee in term of affects, they are directly related
to the nature of the question. An easy question will lead to positive expected affects and a tough question to negative ones.
The values of the virtual recruiter's simulated affects,
$\mathcal{A}_f(t)$ (``f'' stands for felt), is computed using expert rules based on the values
of $\mathcal{A}_d(t)$ and $\mathcal{A}_e(t)$. All these rules are described in \cite{jones2013}.

The list of affects (emotions, moods and attitudes) that are
represented in the model for the virtual recruiter is given in
Table~\ref{RecruiterAffects}. Note that this set is different from the
detected and expected affects. It is based on the literature and on
the mock interview corpus analysis (especially the knowledge
elicitation phases presented in section~3). The emotions are a simple
subset of the OCC model \cite{Ortony1988b} that was selected based on
what practitionners expressed, to different degrees, during the mock
interviews. The moods originated from the ALMA model
\cite{Gebhard2005}, limited to the positive dominance zone (since
recruiters do not show submissive moods in the context of job
interviews) with an additional distinction between bored and
disdaintful, that correspond to two subspaces of (P-,A-,D+) zone that
were distinguished during the knowledge elicitation phase. The
attitudes were identified during the same knowledge elicitation phase and will be discussed in subsection~\ref{sec:attitudes}.

\begin{table}[h!t]
\begin{center}
\begin{normalsize}
\begin{tabular}{|c|c|c|}
\hhline{~|-|-|}
\multicolumn{1}{c|}{}& Positive  & Negative \\ \hline
 & Joy & Distress \\ \cline{2-3}
Emotions & Relief & Disappointment \\ \cline{2-3}
 & Admiration & Anger \\ \cline{2-3}
 & Hope & Fear \\ \hline

 & Relaxed & Hostile \\ \cline{2-3}
 Moods & Exuberant & Bored \\ \cline{2-3}
 &  & Disdainful \\ \hline

 & Friendly & Aggressive \\ \cline{2-3}
 Attitudes & Supportive &  \\ \cline{2-3}
 &  & Dominant \\ \cline{2-3}
 &  Attentive & Inattentive \\ \cline{2-3}
 &  & Gossip \\ \hline

\end{tabular}
\end{normalsize}
\end{center}
\caption{Recruiter affects}
\label{RecruiterAffects}
\end{table}

The computation of moods is based on emotions following ALMA \cite{Gebhard2005}. In the context of our interview simulation, the period is determined by the number of cycle question/answer. Each answer leads to the computation of a new emotions set and these emotions slightly influence the interviewer's mood. The basis for calibration is as follows: after 5 cycles of a specific emotion (anger for example), the virtual recruiter will be in the corresponding mood (hostile). More details can be found in \cite{jones2013}.


\subsubsection{Virtual recruiter's social attitudes}
\label{sec:attitudes}

Several research has shown that one's social attitude is influenced by one's affective state (e.g. his emotions and moods \cite{forgas84}) and the course of the interaction (e.g. the affective reaction of the other \cite{Wegener1994}). For instance in \cite{Wegener1994}, Wegener et al. show that people tend to feel a higher likelyhood toward interlocutors that are in a positive mood.

Relations between attitudes and personality \cite{Snyder1983} and moods and attitudes \cite{Wegener1994,forgas84} has been exhibit in literature. Although we cannot give all the details of these papers here, their results show that one's mood has an influence not only on the interlocutor's attitude, but also on one's own reaction to events. This knowledge has been turned into expert rules that compute values for the attitude of the virtual agent (using the categories elicited in the corpus and presented on table~\ref{RecruiterAffects}). These rules for the computation of attitudes rely on two main parameters: the actual mood of the virtual recruiter (which evolves during the simulation according to agent's emotions) and the initial personality of the virtual recruiter (which will remain in the same state during the simulation). They are presented hereafter.

For the computation of the influence of personality on attitude, based on \cite{Snyder1983}, we propose to use the five model \cite{costa1992} to represent the virtual recruiter's personality using 5 quantitative dimensions (Openness (O), Conscientiousness (C), Extraversion (E), Agreeableness (A) and Neuroticism (N)).


\subsubsection*{Computation of attitudes}

The way we compute attitudes follow this principle: an agent can adopt an attitude according to its personality \cite{Snyder1983} or according to its actual mood \cite{Wegener1994}. For example, an agent with a non-aggressive personality may still show an aggressive attitude if its mood becomes very hostile. The mood compensates the personality and vice versa. For this reason, we use a logical-OR as condition on these two dimensions. As a consequence, in our model, the attitude can be triggered by one of these two dimensions. Then, the maximum value (mood or personality) is kept to compute the corresponding attitude, as is classically done in Fuzzy logics.

The minimum value for a trait to have an influence on the attitude is represented by a simple threshold $\theta$. In our experiments, this threshold is a constant value set to $0.5$; but  it could be a function or a value that evolves in time. The following paragraphs present the computation rules for each attitude.

According to \cite{costa1992}, friendly attitudes are more present in agreeable persons. Moreover, according to \cite{Isbister2006}, positive and aroused moods are associated to the friendly quadrant. For these reasons, we define $val(friendly_f)$ as the combination of $val(A)$ and $val(exub._f)$. If $(val(A) > \theta) \vee (val(exub._f) > \theta)$, then $val(friendly_f) = max(val(exub._f), val(A))$.

According to \cite{Tremblay2005}, aggressive attitudes are more present in non-agreeable and neurotistic personalities and is correlated to the $hostile$ mood. For this reason, if $((val(A) < \theta) \wedge val(N) > \theta ) \vee (val(hostile_f) > \theta)$, then $val(aggr._f) = max(val(hostile_f), val(N), 1-val(A))$.

Similarly, we define dominant attitudes as: if $((val(E) > \theta) \wedge val(N) > \theta ) \vee (val(hostile_f) > \theta)$, then $val(dominant_f) = max(val(hostile_f), val(N), val(E))$.

For the same reasons, we define supportive attitudes as: if $((val(E) > \theta) \wedge val(A) > \theta ) \vee (val(relax_f) > \theta)$, then $val(supportive_f) = max(val(relax_f), val(A), val(E))$.

The attentive and inattentive dominance are simply computed based on the $conscientiousness$ and the $disdainful$ and $relaxed$ mood, based on the definition of these affects:\\
-- If $(val(C) < \theta) \vee (val(disd._f) > \theta)$, then $val(inatt._f) = max(val(disd._f), 1-val(C))$\\
-- If $(val(C) > \theta) \vee (val(relax_f) > \theta)$, then $val(att._f) = max(val(relax_f), val(C))$

Last, $val(gossip_f)$ was defined using $extraversion$ and the $exuberant$ attitude. If $(val(E) > \theta) \vee (val(exub._f) > \theta)$, then $val(gossip_f) = max(val(exub._f), val(E))$.


\subsection{Attitude on interpersonal circumplex} \label{circumplex}

The non-verbal behaviour model of our agent, presented in the next section,
does not work directly with the categories that were identified in the Knowledge elicitation phase of the corpus collection. On the contrary, it makes use of continuous values; it relies on the annotation of corpus which uses the Friendly and Dominant dimensions of
the interpersonal circumplex \cite{Leary1957}.

\begin{figure}[!ht]
\begin{center}
 \includegraphics[width=\linewidth]{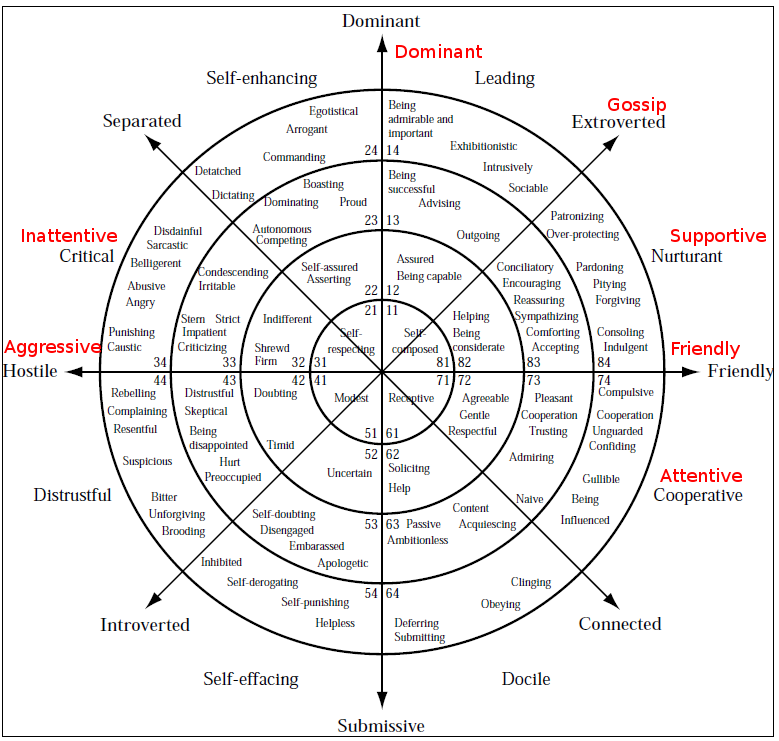}
\caption{Interpersonal circumplex by Isbister}
\label{fig_interp_circump}
\end{center}
\end{figure}

To convert the attitudes represented by categories into continuous values of dimensions, we rely on the work by Isbister \cite{Isbister2006}, summarized in Fig.~\ref{fig_interp_circump}. Our set of attitudes is shown in red on the circumplex, on the line from the center of the circumplex to the outer. The position on this line relies on the intensity of the attitude.


When several attitudes are triggered at the same time, we compute the global attitude (that is the attitude that emerge) as the average of the associated vectors of the various attitudes. The vectors magnitudes influence this average giving more importance to an attitude with a large magnitude (i.e., intensity)). Considering $n$ attitudes $a_i$:\\
\hspace*{5mm}$Friendly = \frac{1}{n}\sum \Big( val(a_i) . Attitude_{Friendly}(a_i) \Big)$\\
\hspace*{5mm}$Dominance = \frac{1}{n}\sum \Big( val(a_i) . Attitude_{Dominance}(a_i) \Big)$

As a result, the affective model gives a level of dominance and friendliness that represents the global attitude of the agent.




\subsection{Evaluation of the affective model}

To evaluate the affective module, we compare the affects that are computed from the affective module with the manual annotation of emotions from the job interview corpus. To perform such an evaluation, we chose arbitrarily one video of this corpus. The affective states of the interviewee and the social attitudes of the recruiter have been manually annotated. Our evaluation consists in comparing if, given the affective states of the interviewee, does the affective module computes social attitudes for the recruiter that are identical (i.e. similar in terms of dimensions representation) as the manually annotated social attitude of the recruiter in the video?

Let us describe briefly the sequence of 8 speaker turns in the video we are using for the evaluation:  in the first 4 questions/answers turns, the interviewee is confident and gives good answers to the recruiter (positive detected affects). The interviewer expectations has been annotated positively during this first sequence. Then, the 4 following questions appear more difficult for the interviewee; he shows expression of negative affects. On the other hand, expectations of the recruiter were annotated positive for questions 5 and 6 and negative for questions 7 and 8.

\begin{figure}[!h]
\begin{center}
 \includegraphics[width=\linewidth]{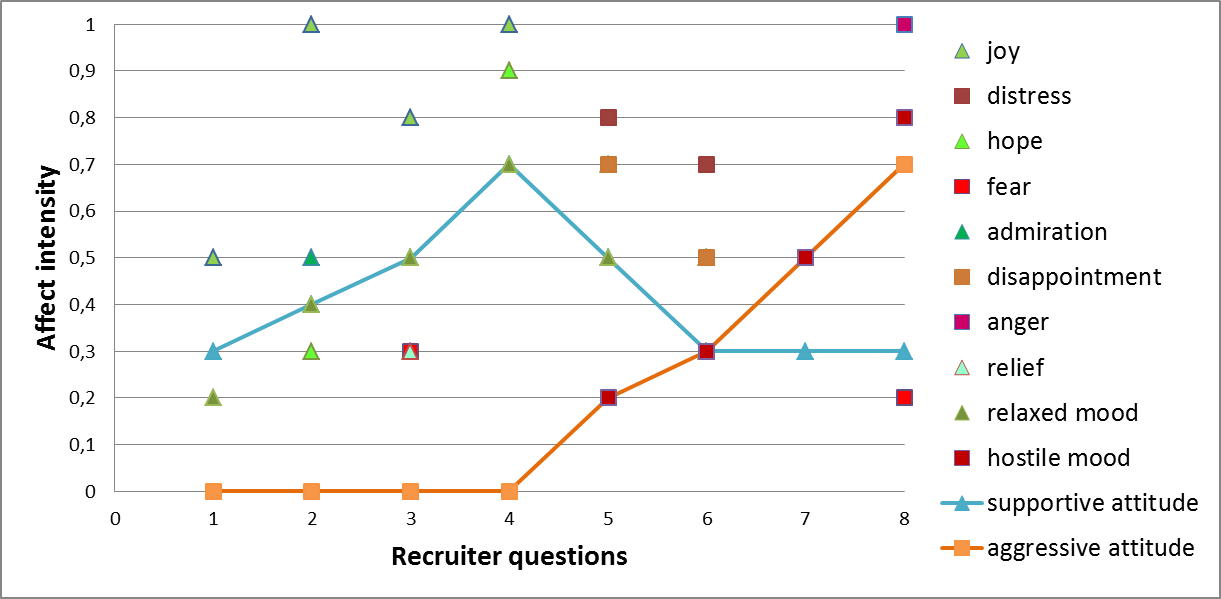}
\caption{Recruiter affects after each question}
\label{fig_result}
\end{center}
\end{figure}

\begin{figure}[!h]
\begin{center}
 \includegraphics[width=\linewidth]{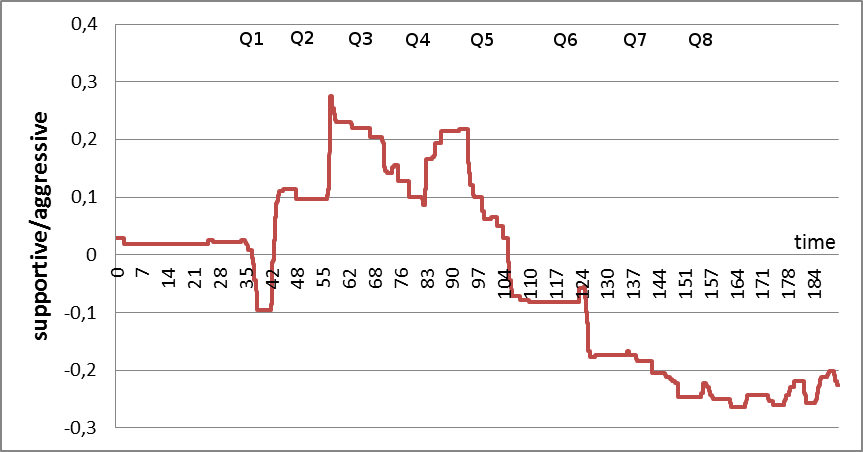}
\caption{Recruiter state in real time}
\label{fig_real}
\end{center}
\end{figure}

Fig. \ref{fig_result} and Fig. \ref{fig_real} show the recruiter affects. Fig. \ref{fig_result} shows the output for the recruiter's affects after each of its question. We can notice that positive affects (triangles) are positively correlated with supportive attitude while negative affects (squares) are positively correlated with aggressive attitude. Fig. \ref{fig_real} displays the annotated social attitudes of the recruiter as it evolves during the interaction. Positive values mean that the recruiter is perceived supportive by the user while negative values as aggressive. We also indicate where the 8 questions happen in the course of the interaction.

Comparing both data, the outputs of the affective module and the manual annotation, we can remark that the results are not really comparable in term of intensity of social attitudes. However the attitudes computed by the affective module coincide with the manually annotated attitudes. The variation of attitudes from supportive to aggressive happens in both cases.
This example shows that our affective module computes, for a given input (the affect states of the interviewee), similar attitudes for the recruiter of those that are perceived in real human-human job interview.

\section{Expression of attitudes}
\label{attitudesDisplay}
Research has shown that non-verbal behaviours convey interpersonal attitude (see Section \ref{Relatedwork}. However it is insufficient to look at signals independently of the other surrounding signals: a smile is a sign of friendliness, but a smile preceded by head and gaze aversion conveys submissiveness \cite{Keltner95}. It is also important to consider how signals from the virtual agent and signals from the user are sequenced (\textit{e.g.} for mimicry).


In our work, to give the capability to a virtual agent to convey attitudes, we choose a corpus-based approach to find how  attitudes are expressed through signal sequences. Rather than finding direct mapping of attitudes to sequences of behaviors, we are interested in characterizing when a sequence of behaviours changes the perception of the associated attitudes. It allows us to capture how attitudes evolve through the interaction through the dynamic display of nonverbal behaviours. Using sequence mining techniques, we extract sequences and characterize how likely they are to convey an attitude variation. The result of this analysis is integrated in the \textit{Behaviour Planner} of the SAIBA architecture, that receives the attitude computed by the \textit{Affective Module}, and chooses a sequence of non-verbal behaviours accordingly. In the following section, we present an algorithm that extract sequences of non-verbal signal.

\subsection{Extraction of sequences from the corpus}
In order to extract significant sequences of non-verbal signals conveying interpersonal attitudes from our corpus, we use a\textit{ frequent sequence mining} technique \cite{Srikant96}. Such techniques have been widely used in protein classification \cite{Ferreira05}, and recent work have used these  techniques in multimodal contexts: in a video game context, Martinez and Yannakakis studied the sequences of user keystrokes, game related events and user physiological signals \cite{Martinez11}. To the best of our knowledge, this technique has not yet been applied to analyse sequences of non-verbal signals. In order to apply the frequent sequence mining technique to our data, we proceed through the following six steps. 

The first step consists of parsing the non-verbal annotations files (coded in the ELAN format), filtering the annotation modalities to investigate and converting every interaction's annotations into a list containing all the non-verbal behaviours in a sequence. 

The second step as for objective to find events to segment the interactions: indeed, frequent sequence mining techniques require a dataset of sequences. In our case, our data consists of 3 continuous interactions (videos of 3 full job interview sessions). Since we investigate which sequences of behaviour convey attitudes, we decide to segment the full interactions with attitude variation events: \textit{attitude variation events} are the timestamps where the attitude begins to vary. To this end, we parse the attitude annotations files, smooth them and find the timestamps where an attitude starts to vary. More details can be found in \cite{Chollet13MMC}. 

Thirdly, we aggregate the attitude variation events into different clusters. Using a K-Means clustering technique, with $k=4$. For friendliness and dominance, we obtain 4 clusters separating the attitude variation events in clusters for large increases, small increases, small decreases or large decreases in value.

The fourth step consists of segmenting the full interaction sequences with the attitude variation events obtained from step 2. Following this procedure, we obtain 245 segments preceding dominance variations and 272 preceding friendliness variations. These two sets are split further depending on which cluster the attitude variation event belongs to. For instance, we have 79 segments leading to a large drop in friendliness, and 45 segments leading to a big increase in friendliness.

Step 5 consists of applying the frequent sequence mining algorithm to each set of segments. We used the commonly used Generalized Sequence Pattern (GSP) frequent sequence mining algorithm described in \cite{Srikant96}. The output of the GSP algorithm is a set of sequences along with their support, \textit{i.e.} the number of times a sequence happens in the dataset.

However, the support is an insufficient measure to analyse how a sequence is characteristic of an attitude variation: for instance, having the gaze move back and forth to the interlocutor happens very regularly in an interaction. Thus it will happen very often before an attitude variation (i.e. it will have a high support), even though it is not sure that it is characteristic of an attitude variation. The objective of step 6 is to compute \textit{quality measures} to assess whether a sequence is really characteristic of an attitude variation. Based on \cite{Tan05}, we choose to compute \textit{confidence }and \textit{lift }quality measures for every sequence. The confidence represents how frequently a sequence is found before an attitude variation, and the lift represents how the sequence occurs before an attitude variation more than expected (the higher the value, the more likely it is that there is dependence between the sequence and the attitude variation).

In Tab.~\ref{extractedseqexamples} we show examples of extracted sequences found using this process, and the corresponding measures. The $Sup$ column corresponds to the support of the sequence and the $Conf$ column to the confidence of the sequence.

\begin{table}[ht]
\begin{center}
\begin{normalsize}
\begin{tabular}{|c|c|c|c|c|}
\hhline{-|-|-|-|-|}
\scriptsize Sequence   & \scriptsize Variation& \scriptsize $Sup$ & \scriptsize $Conf$ & \scriptsize  $Lift$\\ 
 & \scriptsize  type & & & \\ \hline
\scriptsize EyesAt-> EyesAway-> & \scriptsize  Friendliness & \scriptsize  0.027 & \scriptsize  0.727 & \scriptsize  2.72 \\
\scriptsize EyebrowUp->HeadShake  & \scriptsize  Big Decrease  & & & \\ \hline 
\scriptsize HeadAt->GestComm-> & \scriptsize Dominance  & \scriptsize 0.023 & \scriptsize 0.6 & \scriptsize 5.38 \\
\scriptsize EyebrowUp->Smile  & \scriptsize  Big Increase & & & \\ \hline 
\end{tabular}
\end{normalsize}
\end{center}
\caption{Examples of extracted sequences}
\label{extractedseqexamples}
\end{table}

In the next section, we present how appropriate sequences are chosen using \textit{lift }and \textit{confidence }values.


\subsection{Sequences selection for attitude expression}

The attitude expression is handled in our virtual agent platform in the \textit{Behaviour Planner} module. The original role of this module is to instantiate communicative intentions into different signals, such as gestures or facial expressions. In addition to this, we extend the module to have it plan sequence of signals that will convey a chosen attitude. As shown in Fig.~\ref{fig:general architecture}, it receives attitude values computed by the \textit{Affective Module},  in the form of a pair of values, $Friendly$ and $Dominance$. 

\begin{figure}[ht!]
     \begin{center}
        \subfigure[HeadAway]{%
            \label{fig:first}
            \includegraphics[width=0.3\columnwidth]{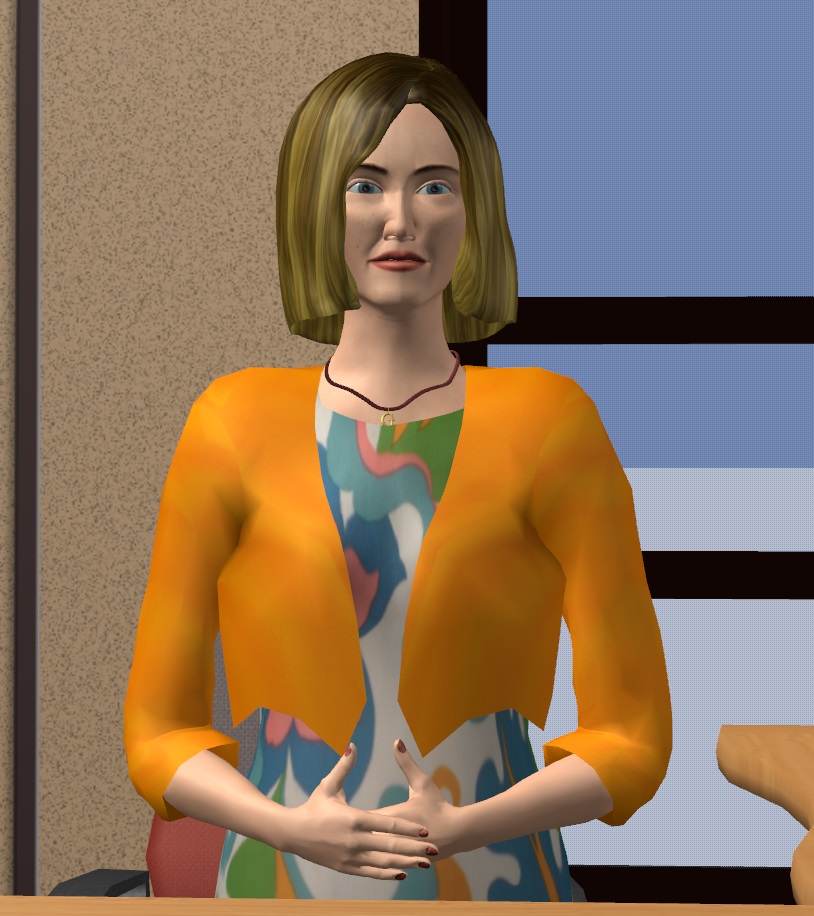}
        }%
        \subfigure[HeadAt]{%
           \label{fig:second}
           \includegraphics[width=0.3\columnwidth]{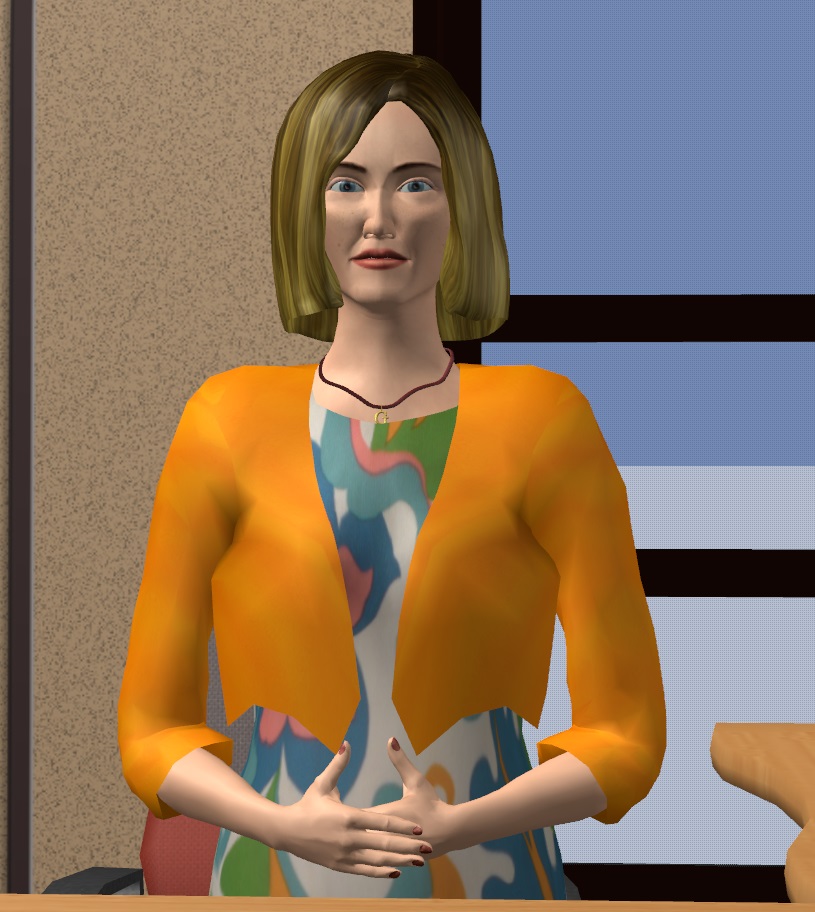}
        } 
        \subfigure[Communicative gesture]{%
            \label{fig:third}
            \includegraphics[width=0.3\columnwidth]{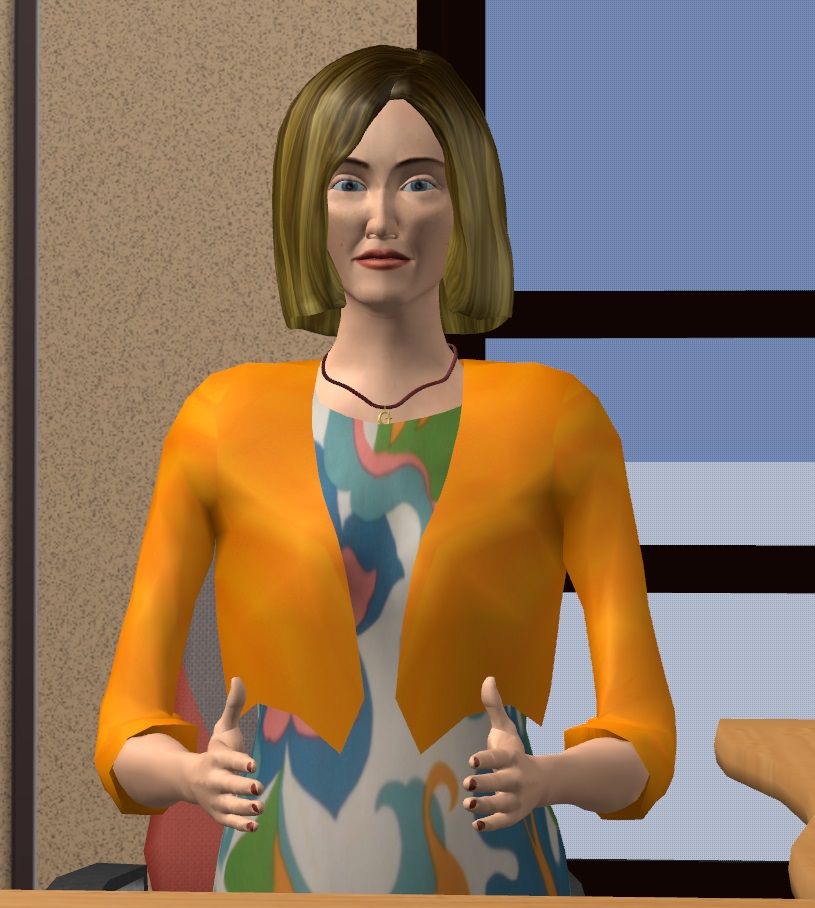}
        }\\%
        \subfigure[Eyebrows raise]{%
            \label{fig:fourth}
            \includegraphics[width=0.3\columnwidth]{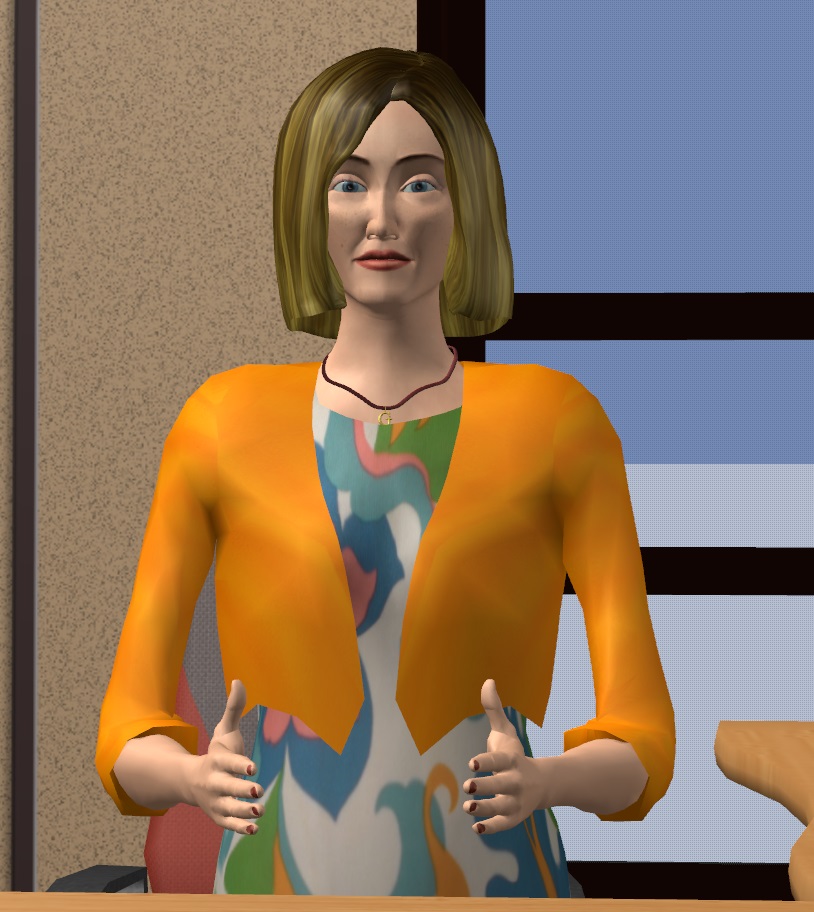}
        }%
        \subfigure[Smile]{%
            \label{fig:fifth}
            \includegraphics[width=0.3\columnwidth]{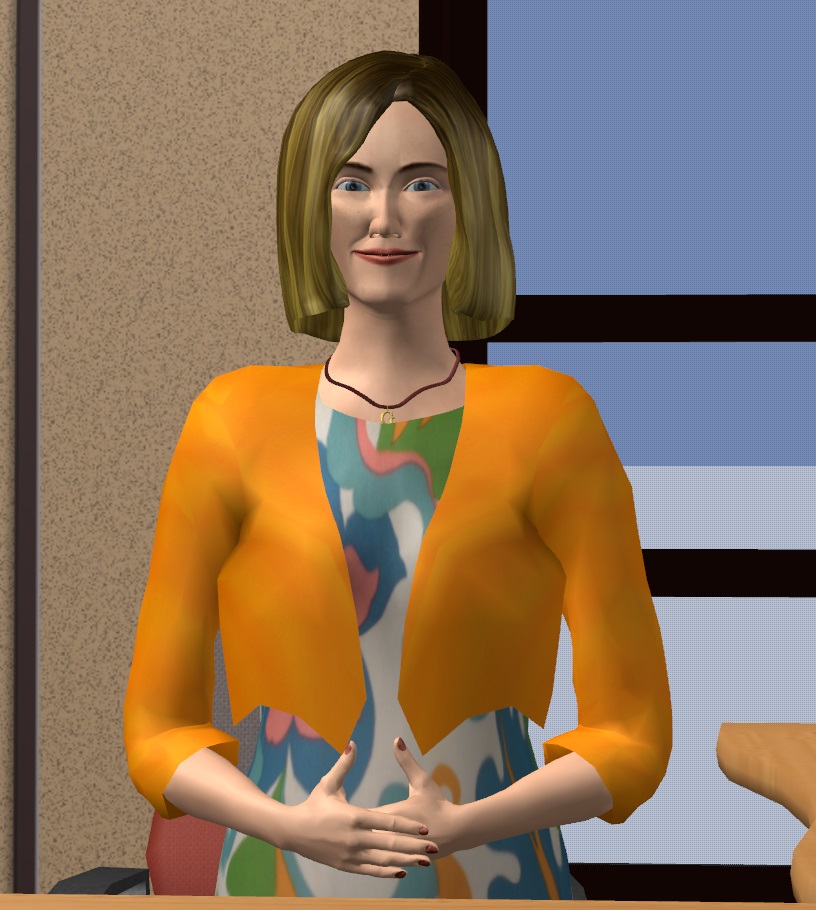}
        }%
        \subfigure[Rest]{%
            \label{fig:sixth}
            \includegraphics[width=0.3\columnwidth]{2_headat.jpg}
        }%
    \end{center}
    \caption{%
        Exemple of the non-verbal sequence $HeadAt->GestComm->EyebrowUp->Smile$ (from Tab.~\ref{extractedseqexamples}) expressed by the virtual agent.
     }%
   \label{fig:gretasequence}
\end{figure}

The \textit{Behaviour Planner} computes $\Delta_{Friendly}(a_n, a_{n-1})$ and $\Delta_{Dominance}(a_n, a_{n-1})$, the differences between the chosen attitude in the last dialogue turn $a_{n-1}$ and the new attitude $a_n$ to express. 
The \textit{Behaviour Planner} then selects the set of sequences corresponding to the attitude variation. Moreover, we look for a non-verbal sequence that will be feasible in the current context. For instance, if the virtual agent is already leaning backwards, the potential sequences containing a change of posture backwards will be discarded. To select the appropriate sequence, for every candidate sequence $seq$ and for target attitude variations $\Delta_{Friendly}$ and $\Delta_{Dominance}$, the following values are computed considering its confidence $Conf$ and lift $Lift$ measures:
$$Val(seq) = Conf(seq,\Delta_{Friendly}) * Conf(seq,\Delta_{Dominance})$$
$$* Lift(seq,\Delta_{Friendly}) * Lift(seq,\Delta_{Dominance}) / \lambda$$

with $\lambda$ being a coefficient used to weigh down sequences if they have been recently displayed, in order to avoid repetition of the same sequences. The highest scoring sequence is chosen by the \textit{Behaviour Planner} and passed on to the \textit{Behaviour Realizer}, in turn producing the animation of the virtual agent. An example of such an animation is shown in Fig.~\ref{fig:gretasequence}.

Our next step, to be performed in the very near future, consists of conducting user perceptive tests to validate that the sequences computed by the \textit{Behaviour Planner} convey the chosen attitude when expressed by the virtual agent.



\section{Concluding remarks}

In the context of social coaching with virtual agents, this paper
proposes a new architecture for expressive agents that can reason
about and display social behaviours. Unlike classical reactive models,
our approach combines an affective reasoner that generates affects for
the virtual character with a sequence selection mechanism based on a
domain corpus annotated on two dimensions: dominance and friendliness.

The methodology we propose contains several stages, from corpus
collection and annotation, knowledge elicitation with experts for the
definition of rules, implementation of behaviours corresponding to 
sequences and sequence selection based on the generated internal
affects. The originality of our model is twofold: first, we rely on an
affective model that receives as inputs expectations on user affects,
compares them to user perceptions and computes agent affects accordingly. All
affects were designed after knowledge elicitation phases with domain
experts. Second, the expression of social attitudes does not only
consider the signals to convey such attitudes but also the non-verbal
behaviours context. Contrary to most existing models of communicative
behaviours for ECAs, in our model, to display an attitude, we generate
a sequence of non-verbal signals taking into account the signals the
agent has shown previously as well as the expressions displayed by its
interlocutors. We used a corpus-based approach to identify sequences
of non-verbal behaviours that display variations in attitudes.

This architecture, whose components have been tested separately, has
been integrated using the SEMAINE platform \cite{Schroder12} and is currently being
tested with real users. This will allow us to validate the global
behaviour of our platform in the context of social coaching. However,
several components can still be improved. One first limit of our model
is that we assume exact inputs from the perception module. Dealing
with imprecision in social signal interpretation is one of our next
steps and we intend to use probabilistic or fuzzy theories. In
addition, we intend to provide the affective reasoner with a
representation of the interaction from the recruiter's point of
view. We believe that allowing the recruiter to reason about the
actual and potential behaviour of the applicant, following a Theory of
Mind paradigm, will allow a more credible decision process. We will also take the sequence extraction procedure further to take into account the user's non-verbal signals. This will require the \textit{Behaviour Planner} sequence selection method to plan for the occurrence of user signals, and to react accordingly when the user does not react as planned.

%

\bibliographystyle{abbrv}
\bibliography{idgei14}

\end{document}